\documentclass[pre,aps,floatfix,nofootinbib,superscriptaddress,two column]{revtex4-2}
\usepackage{graphicx}
\usepackage{dcolumn}
\usepackage{latexsym}
\usepackage{hyperref}
\usepackage{amsmath, amsthm, amssymb}
\usepackage{epsfig}
\usepackage{bm}
\usepackage{geometry}
\usepackage[dvipsnames]{xcolor}
\usepackage{placeins}
\usepackage{diagbox}
\usepackage{relsize}
\begin{document}

\preprint{APS/123-QED}

\title{String Formation and Arrested Ordering Kinetics in Nematics Induced by Polar Particles}% Force line breaks with \\

\author{Pawan Kumar Mishra$^{a,b}$}
% \email[]{pawankumarmishra.rs.phy19@itbhu.ac.in} % Add a common or corresponding author's email if needed
% \email[]{Current affiliation: Mechanobiology Institute, National University of Singapore, 117411 Singapore}
\affiliation{Indian Institute of Technology (BHU), Varanasi, 221005, India}

\author{Partha Sarathi Mondal$^a$}
\email[Corresponding author:]{parthasarathimondal.rs.phy21@itbhu.ac.in}
\affiliation{Indian Institute of Technology (BHU), Varanasi, 221005, India}

\author{Pratikshya Jena}
\affiliation{Indian Institute of Technology (BHU), Varanasi, 221005, India}

\author{Shradha Mishra}
\affiliation{Indian Institute of Technology (BHU), Varanasi, 221005, India}
% \email{pawankumarmishra.rs.phy19@itbhu.ac.in} % Add a common or corresponding author's email if needed
% \affiliation{%
% Department of Physics, Indian Institute of Technology (BHU), Varanasi\\
% %This line break forced with \textbackslash\textbackslash
% }

\date{\today}% It is always \today, today,

\begin{abstract}
Our study explores the mixture of polar particles in apolar environment. We employ a coarse-grained approach to model the mixture,  where polar particles are in minority.The interaction between polar and apolar components is incorporated via a coupling term in the free energy.  Coupling generates local interaction in the system which results in the formation of string like structures connecting a pair of half integer topological defects. The increase in the coupling strength or the density of polar particles results in the: Sharper strings with larger probability of connecting the topological defects of same charge and the enhanced dynamics of topological defects.
However, the ordering kinetics of the system shows the delayed coarsening for larger coupling or polar density. Our results can be used to develop controlled kinetics  as well as to detect the impurities in liquid crystals.
   
\end{abstract}

\maketitle
\def\thefootnote{a}\footnotetext{These authors contributed equally to this work}\def\thefootnote{\arabic{footnote}}
\maketitle
\def\thefootnote{b}\footnotetext{Current affiliation: Mechanobiology Institute, National University of Singapore, 117411 Singapore}\def\thefootnote{\arabic{footnote}}

%\tableofcontents

\section{\label{sec:level1}Introduction}
Polar-apolar mixtures, in which one component exhibits polar symmetry (e.g., fish, bacteria, molecular motors, or certain types of synthetic rods) and the other shows nematic or apolar symmetry (e.g., liquid crystals), are common in both biological and synthetic systems \cite{trivedi2015bacterial,sasaki2014colloidal,tatarkova2007colloidal}. When polar particles are introduced into a nematic liquid crystal environment, fascinating collective behaviors and complex defect dynamics can emerge due to interactions between the polar and apolar phases. Liquid crystals are particularly known for their ability to host topological defects in nematics. In these nematic liquid crystals, topological defects carry charges of \( \pm \frac{1}{2} \) and move diffusively, eventually annihilating as the system relaxes toward equilibrium \cite{de1993physics, Chandrasekhar_1992}. Remarkably, this defect behavior persists even when the system is active---where the medium consists of self-driven units leading to the concept of ``active nematics'' \cite{mishra2014aspects,mishra2014giant,kumar2020active,doostmohammadi2018active,giomi2014defect,thampi2016active,thampi2014instabilities}.\\

The inclusion of foreign particles can significantly alter the dynamics of both active and passive nematic systems. Several studies have examined the effects of bacteria and other inclusions in liquid crystal environments, uncovering the formation of unique structures and dynamics that are absent in pure systems \cite{PhysRevE.108.024701,zhou2017dynamic,genkin2017topological,zhou2014living,mushenheim2014dynamic,turiv2013effect,sokolov2015individual,amiri2022unifying,ngo2012competing}. In this work, we investigate the general case of a polar-apolar mixture, employing a coarse-grained approach to study the interactions between a nematic liquid crystal and dilute polar particles. This framework enables us to understand how passive polar particles affect the structure and dynamics of nematic like topological defects and ordering kinetics within the apolar liquid crystal. This makes our study different from the previous spin-based models, where polar and nematic interactions coexist \cite{lee1985strings,carpenter1989phase,mondal2024ordering}, but in our approach, we treat polar and apolar particles as distinct components interacting via a coupling term.\\
  
In this paper, we introduce a mixed system of apolar and polar particles, where polar particles are less in number. The system is modeled using the time Dependent Ginzburg-Landau (TDGL) equations for the local order parameters for polar and apolar particles, which is obtained by the variation of a combined free energy functional of pure polar, apolar and coupling term. The strength of coupling and density of polar particles are the two key control parameters in the system. Both parameters can be tuned in the experiments by tuning the flexibility and density of polar particles.\\
We study the mixture by assuming external polar particles to be passive and present in small numbers, e.g., a few polar particles present in the liquid crystal. In this framework, the steady state of the pure polar system is disordered, and the average density of the apolar particles is chosen in such a way that its steady state is ordered in the absence of polar particles. \\
We focus our study on the behavior of topological defects formed in the liquid crystal during the evolution of the system, starting from the disordered configuration of both apolar and polar particles. 
The relative strength of the two types of interaction dictates the kinetics and steady state behaviour of the system. When quenched from the random disordered state for both types of particles, the apolar particles try to form the ordered domains, and polar particles prefer the disordered state, but the coupling works as a communicating medium between the two types of particles.\\ 
We found that (i) for the small coupling strength, the interspecies interaction does not have much impact on the dynamics of either species, (ii) beyond a threshold value of coupling, the interaction starts to dominate and string type structures are formed connecting the two-topological defects of same or opposite charges; (iii) The structure of the strings sharpens and the probability of connecting the two ends of the strings with the topological defects of the same charge increases and  (iv) The dynamic of the two types of defects increases with coupling and density of polar particles. Additionally, the ordering kinetics get arrested due to interaction with the defects of the same charges.  

\section{Model and Numerical Methodology}
 We consider a system where few polar particles are immersed in  the liquid crystal. We model the system using coarse-grained free energy functional for the individual systems and then introduced a coupling between polar and apolar particles. For polar particles, the orientation order parameter is represented by the polarization vector field \(\boldsymbol{P} = (P_x, P_y)\), while for apolar (nematic) particles, it is given by the tensor order parameter \(\boldsymbol{Q}\), which is a $2 \times 2$ tensor with two-independent components $Q_{xx}$ and $Q_{xy}$ in two-dimensions, where $x$ and $y$ are the cartesian space  directions.

The free energy functional \(F_{\Psi}\) for individual system is the Landau-Ginzburg Free energy functional given by:

\begin{equation}
F_{\Psi} = \int d\vec{r} \left[ \frac{\alpha_\Psi(\rho_{0})}{2} \Psi^2 + \frac{\beta_\Psi}{4} \Psi^4 + \frac{\kappa_\Psi}{2} (\nabla \Psi)^2\right]
\label{eq:F_Psi}
\end{equation}

where \(\Psi\) represents the order parameters of the system. $\Psi = \boldsymbol{P}$ and $\boldsymbol{Q}$ for the polar and nematic fields respectively. $\Psi^2 = \boldsymbol{P}\cdot\boldsymbol{P}$ and $\boldsymbol{Q}:\boldsymbol{Q}$, $\Psi^4 = (\boldsymbol{P}\cdot \boldsymbol{P})^2$ and $(\boldsymbol{Q}:\boldsymbol{Q})^2$. $(\nabla \Psi)^2 = (\nabla \boldsymbol{P}):(\nabla \boldsymbol{P})$ and $(\nabla \cdot \boldsymbol{Q}) \cdot (\nabla \cdot \boldsymbol{Q})$. The tensor scalar product for two tensors $A$ and $B$ is given by $A : B = A_{lk}B_{kl}$ and the   scalar product of a vector with a tensor $\vec{C} \cdot A = C_{j} A_{ji}$.  
The coefficient \(\alpha_\Psi(\rho_0)\) is responsible for the order-disorder transition and is given by:
$\alpha_\Psi(\rho_0) = \alpha_0 \left(1 - \frac{\rho_0}{\rho_{c}}\right)$, where \(\rho_{c}\) is the critical density, \(\rho_0\) is the average density of either polar (\(\rho_{p_0}\)) or apolar (\(\rho_{a_0}\)) particles, and \(\alpha_0\) is a constant. The system is in the homogeneous ordered phase for \(\alpha_\Psi < 0\) and in the disordered phase for \(\alpha_\Psi > 0\).   The coefficient \(\beta_\Psi > 0\) ensures the stability of the system. The third term accounts for the elastic distortion in the $\boldsymbol{P}$ or $\boldsymbol{Q}$ fields in the equal elastic constant limit \cite{de1993physics}. Here, \(\kappa_{\Psi}\) are the elastic constants, and for convenience, we take \(\kappa_{\Psi} = \kappa = 1\) and \(\beta_{\Psi} = \beta = 1\) for both the fields.

Due to the interaction between the two types of particles, they prefer to align along  the long axis of each other.  Thus, we introduce a coupling term of the form $(\boldsymbol{Q}:\boldsymbol{PP})$ in the free energy of the mixed system. The total free energy of the system is

\begin{equation}
F_m = F_P + F_Q -\gamma \int (\boldsymbol{Q}:\boldsymbol{PP}) d{\boldsymbol{r}},
\label{ch5eq3}
\end{equation}
%$A : B$ is the tensor  product between two tensors $A$ and $B$.
$\gamma$ is the coefficient of the coupling between two types of fields and it is relevant parameter in the model. For $\gamma=0$, the  two fields are uncoupled. 

The evolution of the order parameters $\boldsymbol{P}$ and $\boldsymbol{Q}$ is governed by a generalized dynamical equation derived from variation of the free energy functional \(F_m\) with additional thermal fluctuations introduced through the  stochastic noise term. The equations for the two fields $\boldsymbol{P}$ and $\boldsymbol{Q}$ are:

\begin{equation}
    \frac{\partial \boldsymbol{P}}{\partial t} = [\alpha_P - \beta_{P} \vert\boldsymbol{P}\vert^2]\boldsymbol{P} + \kappa_{P}\nabla^2 \boldsymbol{P} + 2\gamma(\boldsymbol{Q} \cdot \boldsymbol{P})  + \zeta_{P}({\bf r}, t)
    \label{time_evolution_P}
\end{equation}
where, $\vert\boldsymbol{P}\vert^2 = P_x^2 + P_y^2$
and 
\begin{equation}
\begin{aligned}
    \frac{\partial Q_{ij}}{\partial t} = & \; [\alpha_{Q} - \beta_{Q} (\boldsymbol{Q}:\boldsymbol{Q})]Q_{ij} + \kappa_{Q}\nabla^2 Q_{ij} \\
    & + 2\gamma(P_i P_j - \frac{1}{d} \delta_{ij}\vert\boldsymbol{P}\vert^2) + \zeta_{Q}({\bf r}, t)
\end{aligned}
\label{time_evolution_Q}
\end{equation}

 Here, $\zeta_P$, $\zeta_Q$ are the Gaussian white noise terms representing the thermal fluctuations. The noise satisfies the properties \(\langle \zeta_\Psi \rangle = 0\) and \(\langle \zeta_\Psi(r,t) \zeta_\Psi(r',t') \rangle = \eta_{\Psi} \delta(r-r') \delta(t-t')\), where \(\eta_{\Psi} = \eta\) is the noise strength for both the fields.

We study the system in two dimensions. The dynamics of both the polarization $\boldsymbol{P}$ and the tensorial order parameter $\boldsymbol{Q}$ are analyzed. The intrinsic time  \(\tau = \frac{1}{\alpha_0} = 1\) and  length \(l_0 = \sqrt{\kappa \tau}\) scales are used to rescale the system. The equations are made dimensionless using $\tau$ and $l_0$. The details of numerical simulation are provided in  Appendix.\ref{app:rescaled equation}.\\

\begin{figure*}
    \centering
    \includegraphics[width=\textwidth]{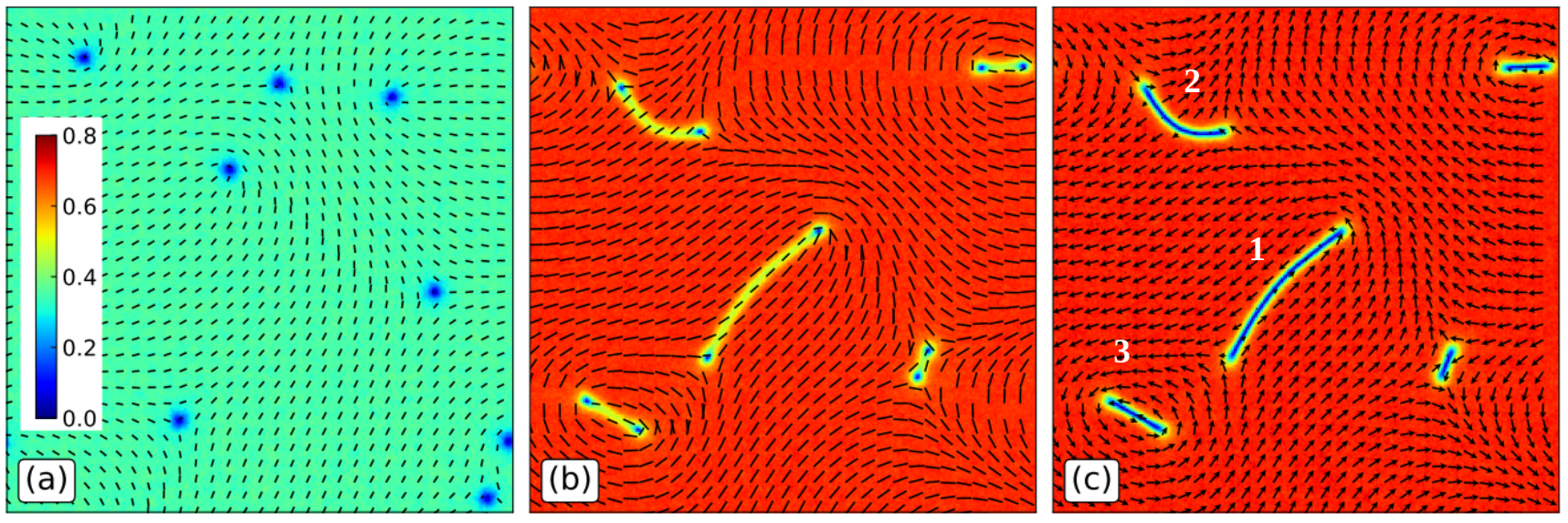}
   
 \caption{Visualization of order parameters. Panels (a) and (b) illustrate the $\boldsymbol{Q}$ field for $\gamma = 0$ and $\gamma = 1.0$, respectively, while panel (c) depicts the $\boldsymbol{P}$ field for $\gamma = 1.0$ for $\overline{\rho}_{p_0} = 0.10$. In panel (a) and (b), the lines represent the direction of the local nematic director, while in panel (c), the arrows represent the local direction of $\boldsymbol{P}$-field.}

    \label{fig1}
\end{figure*}

The system is initialized from a randomized and homogeneous state of apolar particles with an average apolar density \(\rho_{a_0} = 0.75\), and critical densities \(\rho_{pc} = \rho_{Qc} = 0.5\). The value of \(\rho_{a_0}\) is chosen such that the apolar particles are in the ordered phase. The strength of the noise \(\eta\) is set to \(0.1\). The rescaled average density of polar particles $\overline{\rho}_{p_0}=\frac{\rho_{p_0}}{\rho_{pc}}$ and the coupling strength \(\gamma\) are the primary control parameters in the study. \(\rho_{p_0}\) is varied to ensure \(\rho_{p_0} \ll \rho_{pc}\) (i.e. $\overline{\rho}_{p_0}\ll 1.0$),  the pure polar particles in the disordered phase, while \(\gamma\) is varied in the range \([0, 1]\). The focus of our study is different from the other coupled polar-apolar mixtures, where the density of both types of particles are comparable \cite{PhysRevE.108.024701, vats2024surface} as well as the systems are studied in confinement \cite{vafa2025phase}. 

%By systematically tuning \(\rho_{p_0}\) and \(\gamma\), we investigate the influence of these parameters on the interaction between polar particles and the nematic field, particularly focusing on the role of the coupling strength in determining the emergent structures and dynamics.

\section{Results}
 In this study, we examine how coupling influences the structure formation and dynamics within the system. Our focus lies in observing the time evolution of the nematic order parameter field $\boldsymbol{Q}$, and the polarization vector field $\boldsymbol{P}$ as governed by Eq. (\ref{time_evolution_Q}) and (\ref{time_evolution_P}) respectively. The kinetics of the system very much depends on the formation of oppositely charged disclinations, which are topological defects of charge half-integer and integer in pure apolar and polar systems, respectively.\\

{\textbf{\em Strings in the system:}} We begin by illustrating the effect of coupling on both polar and nematic fields through snapshots in FIG.\ref{fig1}. Panels (a) and (b) show the  $\boldsymbol{Q}$-field, for $\gamma = 0$ and $\gamma = 1$, respectively, while panel (c) displays the $\boldsymbol{P}$-field. The color represents the magnitude of two fields in respective snapshots. In panels (a) and (b), the lines represent the local nematic director, while in panel (c), the arrows show the local direction of the $\boldsymbol{P}$-field. At zero coupling ($\gamma = 0$), the polar species remains disordered, and the nematic system exhibits typical behavior for liquid crystals: the $\boldsymbol{Q}$-field exhibits half-integer defects which diffuse freely and annihilate upon colliding with opposite-charge defects. 
As the coupling is turned on, the polar species begin to align, and the direction of local polar ordering is influenced by the local nematic ordering and {\em vice versa}. At a finite $\gamma$, small domains of $\boldsymbol{P}$-field start forming and on further increase in coupling strength, new structures, which we call `strings', in $\boldsymbol{Q}$ and $\boldsymbol{P}$ fields appear in the system, connecting two topological defects. 

\begin{figure*} 
         \includegraphics[width=0.99\textwidth]{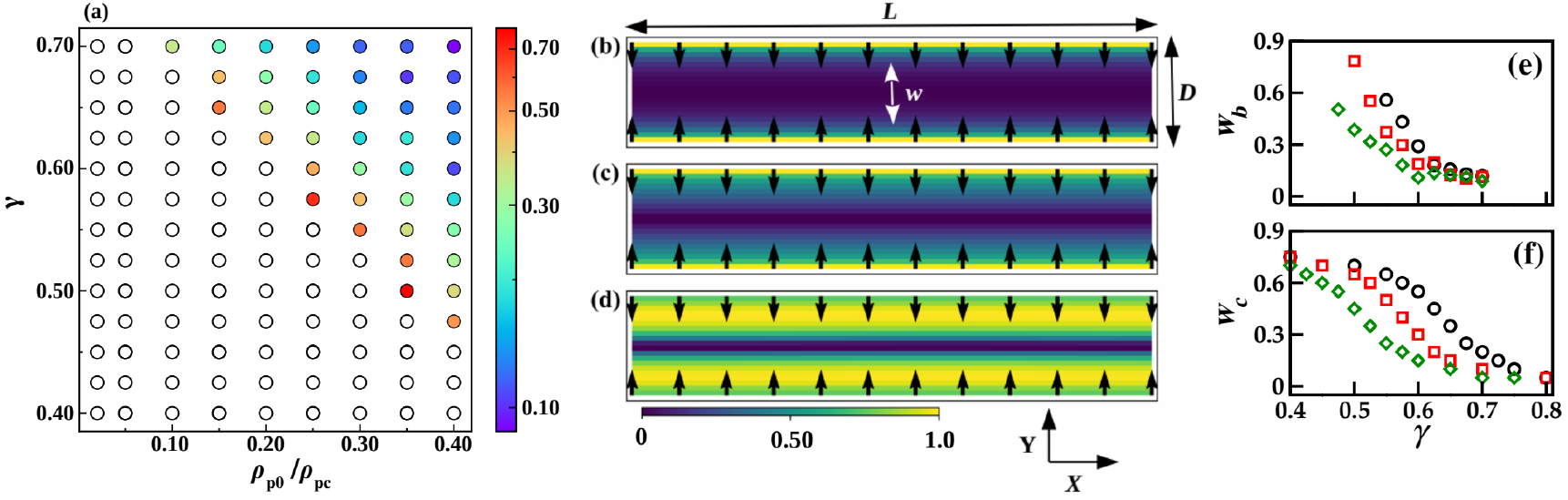}
 
   \caption{(a) Phase diagram illustrates the  width of strings $W_b$ (color map)  in the ($\overline{\rho}_{p_0}$-$\gamma$) plane.  The empty symbols indicate the values of $\overline{\rho}_{p_0}$ and $\gamma$ where strings are absent. (b)-(d) Snapshots of the $\boldsymbol{P}$ field for $\gamma = 0$, $0.6$, and $1.0$ in a confined system with perpendicular anchoring of $\boldsymbol{P}$ field at the top and bottom boundaries. The color map represents the magnitude of $\boldsymbol{P}$ field. (e) Plot of $W_b$ versus  \(\gamma\) for three values of  $\overline{\rho}_{p_0}$ $= 0.30$, $0.35$, and $0.40$ represented by $\circ$ , $\square$ and $\diamond$ respectively. (f) Plot of \(W_c = \frac{W}{D}\) versus  \(\gamma\) for three values of the $\overline{\rho}_{p_0}$ is the same as in plot (e). }
\label{fig:width}
\end{figure*}

For the nematic species, the magnitude of $\boldsymbol{Q}$ 
%i.e. $ \lvert \lvert \boldsymbol{Q}  \rvert \rvert=0.5*({Qxx}^2+{Qxy}^2)$ 
along the string is significantly less than that in the surrounding regions, as shown in FIG.\ref{fig1}(b). Similar string-like structures connecting two half-integer topological defects have been reported in recent experimental \cite{ruider2024topological} and theoretical \cite{vats2024surface,vafa2025phase} studies. However, unlike these works, the strings in our case are observed to connect defects of both the same (e.g., two $+\frac{1}{2}$ or two $-\frac{1}{2}$ as marked by `2' and `3' in FIG.\ref{fig1}(c)) as well as of opposite types (e.g., a $+\frac{1}{2}$ and a $-\frac{1}{2}$ as marked by `1' in FIG.\ref{fig1}c)). The  magnitude of $\boldsymbol{P}$ i.e. $\lvert \boldsymbol{P} \rvert=({Px}^2+{Py}^2)$ approaches $0$ along the string with the opposite orientation of $\boldsymbol{P}$-field on its two sides, as shown in FIG.\ref{fig1}(c).
The formation of string can be explained in the following manner: when a defect appears in the apolar field, a strong coupling to the nematic direction leads the drop in polar order, as the elastic free energy penalizes misalignment. However, the corresponding drop in nematic order results from the decrease in polar order. Hence, the coupling term is sensitive not only to alignment but also to the magnitudes of the respective order parameters. It leads to the almost disorderly arrangement of polar particles with a pair of defects at the two ends of it. Since the nematic free energy prefers the apolar alignment, this encourages the orientation of polar particles antiparallel at the two walls of the string, which further makes a string act like a domain wall for the polar alignment although the formation of the sting is a consequence of the underlying alignment behaviour.\\
Both the fields ($\boldsymbol{Q}$ and  $\boldsymbol{P}$) show the nearly parallel or perpendicular anchoring along the string as shown by the strings labeled as `1' and `2' respectively in Fig.\ref{fig1}(c). We also checked the effect of a large density of polar particles $\overline{\rho}_{p0}$ or high coupling $\gamma$. For large coupling $\gamma >1$ and high density $\overline{\rho}_{p0} \sim 0.8$, the magnitude of local ordering increases, and strings are more probable to be formed between a pair of $1/2$-integer defects of the same charge. However, even for high coupling and large density, we never found integer defects in the system. The snapshots of the $\boldsymbol{Q}$ and  $\boldsymbol{P}$ is shown in FIG.\ref{high_density_high_coup_snaps} in Appendix.\ref{app:hdhc}.\\
Starting from a random state, it takes sometime for the defects to appear and the strings to form. After this initial state, the width of the strings remains the same as long as the defects survive, which we refer to as the dynamical steady state. The presence of strings introduces complexity into the system's dynamics as their configuration evolves with the motion of defects.  The annihilation of defects can either lead to the disappearance of existing strings or the formation of new ones. The animations showing the time evolution of $\boldsymbol{Q}$ and $\boldsymbol{P}$ field provided in Appendix.\ref{movie}.\\
The range of $\gamma$ values for which the strings are formed depends strongly on the mean density of the polar species ($\overline{\rho}_{p_0}$). This dependence is visually represented by a series of snapshots provided in FIG.\ref{phase_diagram}. For larger $\overline{\rho}_{p_0}$, the string formation is observed at relatively lower coupling strengths. 
We also calculate the normalized width $W_b$ of the strings. The details of the calculation of the width are presented in Appendix.\ref{app:Width calculation}. The result obtained for the magnitude of $W_b$ is presented in the form of a phase diagram in FIG.\ref{fig:width}(a) in the $(\overline{\rho}_{p_0}-\gamma)$ plane.We again show $W_b$ {\em vs.}  $\gamma$ for three different values of $\overline{\rho}_{p_0}$ in FIG.\ref{fig:width}(e).  Higher densities of polar particles and stronger coupling strengths lead to narrower strings. In addition to strings connecting defect pairs, we also observe the formation of branched strings connecting multiple defects, as well as closed-loop structures, clearly visible in Fig.\ref{phase_diagram}. The corresponding configurations of the $\boldsymbol{P}$-field for these structures are shown in Fig.\ref{brunched_strings}. These branched strings and loops predominantly appear at early times within the dynamic steady state.
%This finding underscores the significant role that these parameters play in determining the structural and dynamical properties of the system.

The mechanism of formation of the strings involves the interplay of the two length scales: \(\xi_P = \sqrt{\frac{1}{\alpha_{P0}}}\), which governs the variation of the polarization modulus, and \(\xi = \frac{1}{\gamma}\), which controls the variation of the polarization direction under the assumption of a constant nematic field. In practice, the larger of these two length scales dominates. 
To delve deeper into the interplay of two length scales, we switch off the noise and study geometries with rigid boundary condition in one direction, which can generate a wall. We introduced two rigid walls at the top and bottom, where the $\boldsymbol{Q}$-field is fixed with perpendicular anchoring. The $\boldsymbol{P}$-field also has perpendicular anchoring but with opposite orientations at the top and bottom walls. The $\boldsymbol{Q}$-field inside the geometry is kept homogeneously ordered, matching with the walls, and $\boldsymbol{P}$ is random. The periodic boundary condition is used in $X$-direction, and the width of the system $D$ is much smaller than its length $L$. 
%The width of the wall can be compared with the two length scales $\xi_P$ and $\xi$. 
For the regime where \(\xi_P, \xi > D\), no wall is observed. However, when \(\xi_P, \xi < D\), a wall forms. For this geometry and boundary conditions the steady state has a stable wall in the middle of the system ($y = D/2$). In FIG.\ref{fig:width}(b-d) we show the snapshot of the system with magnitude of local $\boldsymbol{P}$ and its direction at the two boundaries for three different $\gamma$ values, $\gamma = 0, 0.6, 1.0$, at a fixed $\overline{\rho}_{p_0}$ for system dimensions $D=20l_0$, and $L=100l_0$. We quantify the width of the wall, $W$, by counting the number of sites where the magnitude of the polar order parameter satisfies \(\lvert  \boldsymbol{P} \rvert < 0.2\) along the $Y$-directions and averaged over $X$-direction at the dynamical steady state. In FIG.\ref{fig:width}(f) we show the plot of relative width of the wall $W_c=W/D$ {\em vs.} $\gamma$ for three different $\overline{\rho}_{p_0}$. Both by increasing $\gamma$ and $\overline{\rho}_{p_0}$, $W_c$ monotonically decreases for a confined system, which is consistent with the plot shown in FIG.\ref{fig:width}(e) for the bulk system.

{\textbf{{\em Kinetics of the system}:}  Now, we shift our focus to examine the effect of coupling on the ordering kinetics of $\boldsymbol{Q}$-field. Starting from a random homogeneous state, the system evolves through the growth of ordered domains of the nematic order parameter field. To characterize the domain growth,  we calculate the correlation length $L(t)$ from the $0.5$ crossing of the two-point correlation function, $C(r,t)$ (see Appendix.\ref{Kinetics of apolar order parameter}). For all values of $\gamma$, $L(t)$ overlaps at initial times. At intermediate times, the $L(t)$ vs $t$ plot exhibits a plateau for $\gamma \geq 0.6$, whereas at late times, the behavior is consistent with $t^{1/2}$ for all $\gamma$, with the magnitude of $L(t)$ strongly dependent on $\gamma$.\\

During the initial times, the nematic system starts growing from random initial conditions, and small domains of  $\boldsymbol{Q}$ begin to emerge. Hence, the growth of $L(t)$ in this regime is independent of $\gamma$, leading to the overlapping curves. At intermediate times, the strings are absent for $\gamma \leq 0.5$, and $L(t)$ matches well with the pure system. For $\gamma \geq 0.6$, the system reaches a dynamical steady state, and the presence of strings affects the growth of $L(t)$, resulting in the plateau observed during the intermediate times. The appearance of the plateau signifies arrested coarsening during intermediate times. The plateau region consistently flattens with increasing $\gamma$, while the length of the plateau region shows non-monotonic behavior: it first increases for $0.6 \leq \gamma \leq 0.8$ and then decreases for $\gamma \geq 0.8$.\\

 At higher $\gamma$ ($\gamma \gtrsim 0.6$), the effect of coupling on the dynamics of the $\boldsymbol{Q}$-field is two-fold:
(i) Increasing coupling leads to the enhancement of the probability of formation of strings connecting defects of the same type, denoted by $P_s$;
(ii) The dynamics of an isolated defect increases with increasing coupling strength. In FIG.\ref{fig3}(c), we show the plot of $P_s$ versus $\gamma$ for three different values of $\overline{\rho}_{p_0}$. The probability $P_s$ increases monotonically with $\gamma$ and $\overline{\rho}_{p_0}$, i.e., with increasing coupling strength, the strings are more likely to connect defects of the same sign.
On the other hand,  to understand the effect of coupling strength on the dynamics of defects, we introduce a pair of artificial half-integer topological defects in the system and its details are presented in the Appendix.\ref{Artificial Defects in Coupled System}. As shown in FIG.\ref{fig3}(d), the speed of defect remains unaffected up to $\gamma \approx 0.6$ in the absence of strings while it monotonically increases as strings appear in the system for $\gamma \gtrsim 0.6$.\\

We propose the mechanism behind the appearance of the plateau and its dependence on coupling through the interplay of two counteracting effects of coupling discussed above. First, as shown in FIG~\ref{fig3}(d), increasing $\gamma$ enhances the dynamics of isolated defects, leading to faster string contraction. This causes quicker annihilation of defects with opposite charges, as illustrated in FIG~\ref{fig_artificial}, promoting faster domain growth for $\gamma = 0.6$ and $0.7$. However, for higher coupling strengths ($\gamma \gtrsim 0.8$), the increased probability of forming strings connecting defects of the same charge, $P_s$, alters this behavior. These strings shrink due to string tension, but as they shorten, the repulsive interaction between like-charged defects becomes dominant, preventing further contraction. Since defects of the same charge cannot annihilate, these short strings persist in the system for extended periods, as shown in FIG~6(d). This persistence leads to arrested kinetics, causing the plateau in $L(t)$ in the intermediate time regime.
To further understand the slower growth kinetics in this regime, we analyze the stress in the system. The local stress is defined as $\sigma_{\text{loc}} = \nabla Q_{xy}$, and the mean stress is calculated as $\sigma = \sum_{\mathbf{r}} |\sigma_{\text{loc}}(\mathbf{r})|$. The time evolution of $\sigma$ for different $\gamma$ values is shown in Figure~\ref{fig3}(b). For larger $\gamma$, the mean stress is higher, and its decay exhibits a plateau in the same time regime as the plateau in $L(t)$. Additionally, at higher coupling strengths, a strong stress peak appears before the plateau, signaling the onset of slower kinetics. This slower stress relaxation occurs because defects induce distortions in the Q-field, generating non-zero stress. The persistent strings connecting like-charged defects hinder defect annihilation, delaying stress relaxation and further slowing domain growth in the intermediate time regime.

\begin{figure}
%  \centering
  \includegraphics[width=0.48\textwidth]{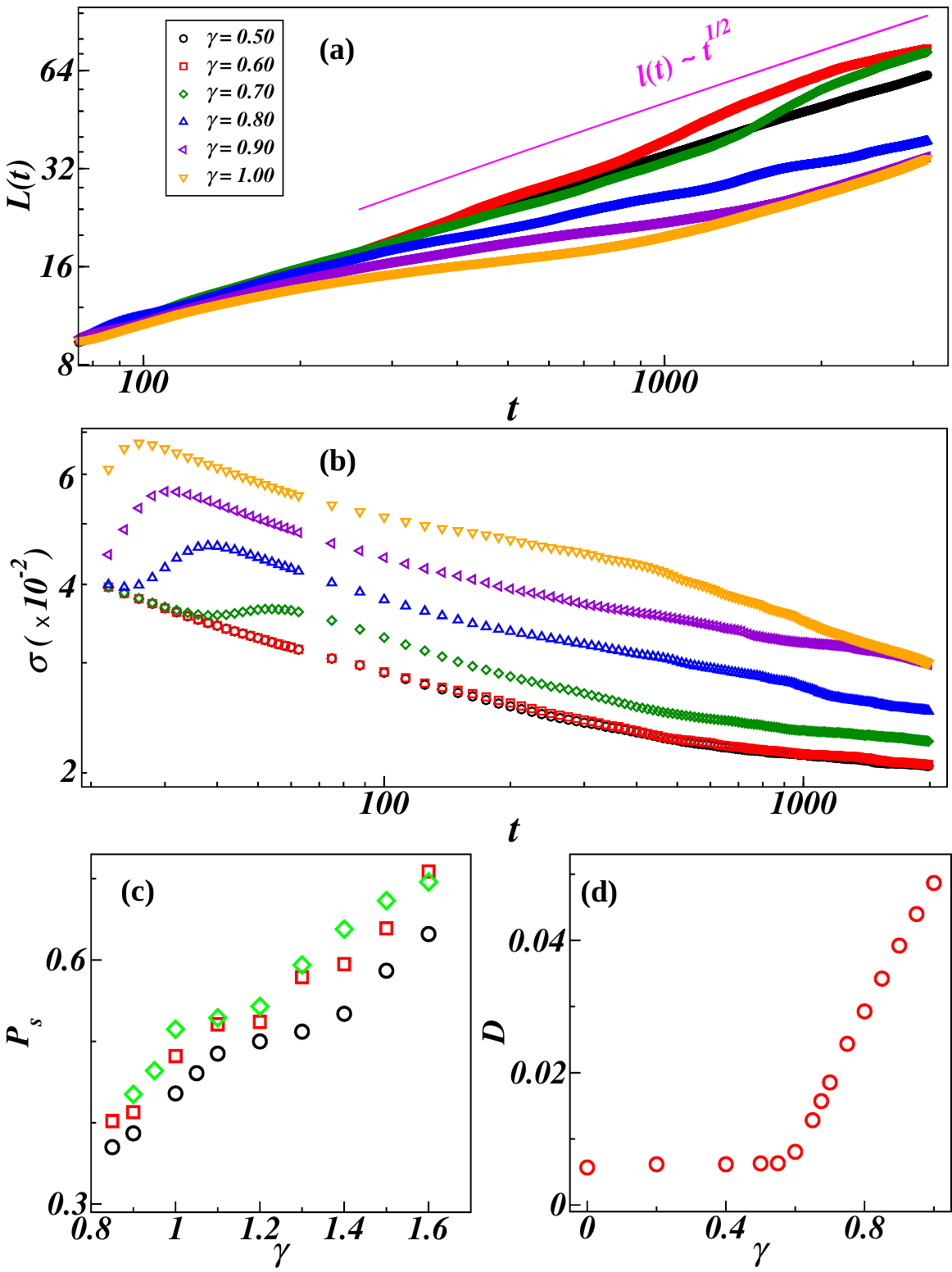}
   \caption{(a) Correlation length $L(t)$ {\em vs.} $t$ for various  $\gamma$ for  $L=1024$ and $\overline{\rho}_{p_0} =0.1$, (b) Mean stress in the system, $\sigma$ {\em vs.} $t$ for $L=1024$ and $\overline{\rho}_{p_0} =0.1$, (c) The probability $P_s$ {\em vs.} $\gamma$ for different $\overline{\rho}_{p_0}$ $= 0.30$, $0.35$, and $0.40$ represented by $\circ$ , $\square$ and $\diamond$ respectively for $L=256$. (d) Speed of the isolated half integer defects ($\pm 1/2$) $v$ {\em vs. } $\gamma$.}
   %(details of the calculation is given in Sec.V in the SM).}

    \label{fig3}
\end{figure}

\section{Conclusion}

%\begin{itemize}
 %{\em Conclusions:-} 
 We study the effect of polar impurities in the liquid crystal system by tuning the coupling strength \(\gamma\) and average density of polar particles \(\rho_{p_0}\). We use the coupled time dependent Ginzburg Landau (TDGL)  equations for order parameters  $\boldsymbol{Q}$ and  $\boldsymbol{P}$. The coupling leads to the formation of string like structures in the nematic order parameter field. The two endpoints of the strings are topological half integer defects. The strings are observed above some optimal values of coupling and density of polar particles.  For polar particles, the magnitude of the order parameter is close to zero along the string, acting as a domain wall separating oppositely oriented domains. The strings are more probable to be connected by the defects of the same sign for larger coupling and density, this results in delayed kinetics of domain formation for strong coupling. 
 Opposite to this effect, the dynamics of single defects is enhanced by the strengthening of the string.
These results highlight the dual role of $\boldsymbol{Q}-\boldsymbol{P}$ coupling in the system: while moderate coupling accelerates defect annihilation and promotes faster ordering, large coupling stabilizes same-type defect connections, ultimately hindering the ordering process. Additionally, increase in $P_s$ with an increase of $\gamma$ leads to the formation of defect structures similar to $\pm 1$ like topological charge formed by the very small string connecting the two $+ 1/2$ (or two $- 1/2$) defects as shown in FIG.\ref{fig1}(c) for $\gamma=1$. This is counterintuitive since $\pm 1$ charged defects are not stable topological structures for nematic.\\

String-like structures in the nematic phase have also been reported in systems such as confined nematic liquid crystals \cite{meng2025emergentdimermodeltopologicalorder}, ferroelectric nematics \cite{sebastian2020ferroelectric,ma2024half} and nematics under magnetic fields \cite{james2011phase}.  In particular, in Ref.\cite{james2011phase} the authors observed string like structures connecting topological defects of the same sign, in a nematic system coupled to a magnetic field. Notably, the coupling in that system is non-reciprocal, and tuning this non-reciprocity is known to unveil distinct behaviors. In contrast, our system features purely reciprocal interactions, where such strings emerge only transiently during the ordering kinetics. A related observation was made in Ref.\cite{vafa2025phase}, where string-like structures persist into the steady state; however, the dynamics in that case is significantly influenced by confinement and boundary conditions.\\
    
In summary, the current study provides a comprehensive analysis of the behavior of topological defects and string formation in coupled nematic and polar systems. The findings of this study hold significance since it provides a method of controlling the prominence of the strings and, hence, the growth kinetics of the nematic by tuning the density of the polar species and the coupling strength.\\

Our results motivate experimental investigation of analogous systems to assess the broader applicability of our findings. A prominent example is ferronematics \cite{mertelj2013ferromagnetism,mertelj2017ferromagnetic}, where magnetic nanoparticles are dispersed in a liquid crystal host. These systems have been extensively studied and are known for their utility in various technological applications. The current framework can be extended to incorporate activity in the polar species, as realized in living liquid crystals \cite{zhou2014living,sokolov2015individual,genkin2017topological}, where the intrinsic activity leads to unique emergent behaviors. These active systems have shown promise in applications such as bio-sensing and targeted transport. Moreover, the framework is versatile enough to accommodate activity in the background nematic medium itself. Previous studies have shown that tuning the activity and density of the polar component enables control over the dynamics of active nematics \cite{sampat2021polar,mondal2025dynamical}. Overall, the present study not only advances the fundamental understanding of interspecies interactions in complex fluids but also offers a flexible and generalizable model with relevance to both future studies and potential applications. \\

\begin{acknowledgments}
P.K.M. and S. M. thank Julia Yeomans for useful discussions. The authors thank Jacques Prost for useful suggestions, discussions and careful reading of the manuscript. P.K.M., P.S.M. gratefully acknowledge UGC for research fellowship and P.J. acknowledge the DST INSPIRE fellowship  for
funding this project. The support and the resources provided by PARAM Shivay Facility under the National Supercomputing Mission, Government of India at the Indian Institute of Technology, Varanasi are gratefully acknowledged by all authors. S.M. thanks DST-SERB India, ECR/2017/000659, CRG/2021/006945 and MTR/2021/000438  for financial support. P.K.M, P.S.M, P.J. and S.M. also thank the Centre for Computing and Information Services at IIT (BHU), Varanasi.
\end{acknowledgments}

\section*{Data availability statement}
All the data that support the findings of this study are included within the article (and any supplementary files).\\

\onecolumngrid
\appendix
%%%%%%....Simulation Details
\section{Simulation details}\label{app:rescaled equation}
We have used the Euler integration scheme on a square lattice of size \(L \times L\) with periodic boundary conditions in both directions. The Euler discretization scheme is applied with mesh sizes \(\Delta x = \Delta y = 0.5l_0 = 0.5\) and time step \(\Delta t = 0.01\tau = 0.01\). The spatial discretization ensures that coarsening interfaces are treated smoothly, while the temporal discretization is selected to maintain the stability of the numerical scheme. Once all grid points are updated, then it is counted as one simulation step. One simulation step covers  \(\Delta t \) time. The results are obtained for system sizes $256 \times 256$ and $1024 \times 1024$ for time steps $1 \times 10^5$ and  $3 \times 10^5$, respectively. For better statistics, the observables are averaged over $40-50$ independent realizations. The parameters for the system with artificial defects and confined systems are mentioned at respective locations.

%%%%%%....Details of Width Calculation
\section{Width calculation}\label{app:Width calculation}
We quantify the width of the structures connecting the topological defects shown in FIG.1(b) in the main text. To compute the width, we analyze the $\boldsymbol{P}$ field in FIG.1(c) in the main text, which provides better contrast compared to the $\boldsymbol{Q}$-field. The width is determined by counting the number of sites, $n_w$, where the magnitude of the $\boldsymbol{P}$-field in FIG.~\ref{phase_diagram} satisfies \(\lvert \boldsymbol{P} \rvert < 0.2\) along both the \emph{X}- and \emph{Y}-directions, and then calculating the average of these values. To ensure accuracy, this averaging is performed across multiple locations in the system. Structures are considered as strings only if \(n_w \leq 30\). To normalize the width, \(n_w\) is divided by 30, yielding the normalized width \(W_b\). FIG.~\ref{phase_diagram} presents a phase diagram showcasing snapshots of the $\boldsymbol{P}$-field, illustrating variations in string width as a function of different $\overline{\rho}_{p_0}$ and $\gamma$ values.

\begin{figure*}[hbt]
    \centering
    \includegraphics[width=\textwidth]{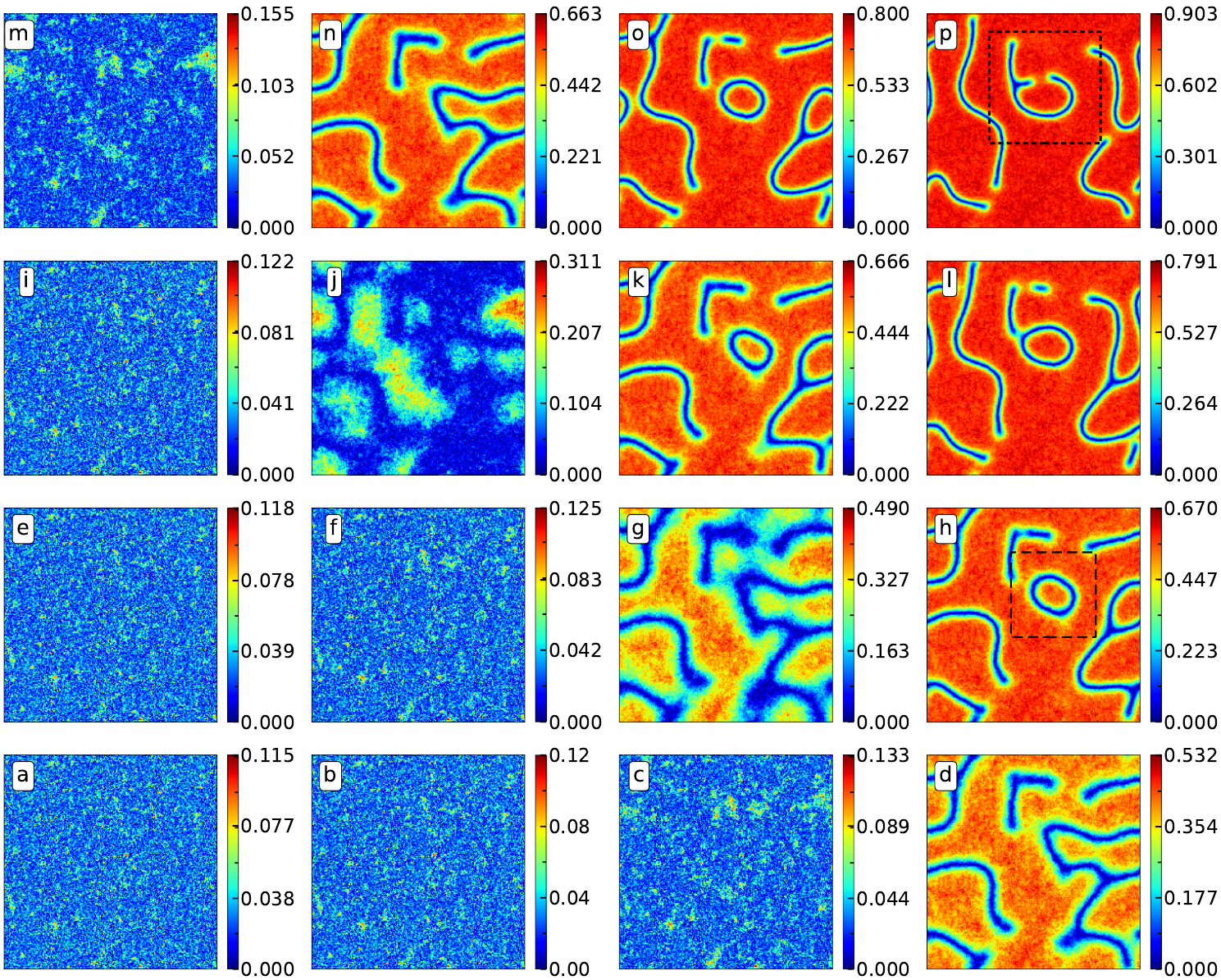}
    \caption{Snapshot of the magnitude of polar order parameter at a fixed time, showing variations in the $\overline{\rho}_{p_0}$ from left to right  and the coupling strength \( \gamma \) from bottom to top. Each row (a-d, e-h, i-l and m-p) represents a different \( \gamma \), varying from weak to strong coupling (\( \gamma = 0.50, 0.55, 0.60, 0.65 \)) respectively. Each column (a-m, b-n, c-o and d-p) represents a different $\overline{\rho}_{p_0}$ varying from low to high (0.10, 0.20, 0.30, 0.40) respectively. The heatmaps in each row illustrate variations of the polar order parameter for the corresponding parameters. The region highlighted by the black box in panel (h) corresponds to a loop configuration, while that in panel (p) depicts a branched string structure. The corresponding orientation fields of the polar species for the loop and branched string configurations are shown in Fig.~\ref{brunched_strings}(a) and (b), respectively, where we present a zoomed in view of the region inside the box.}
    \label{phase_diagram}
\end{figure*}

% ......brunch strigs and circular structures
\begin{figure*}[hbt]
    \centering
    \includegraphics[width=\textwidth]{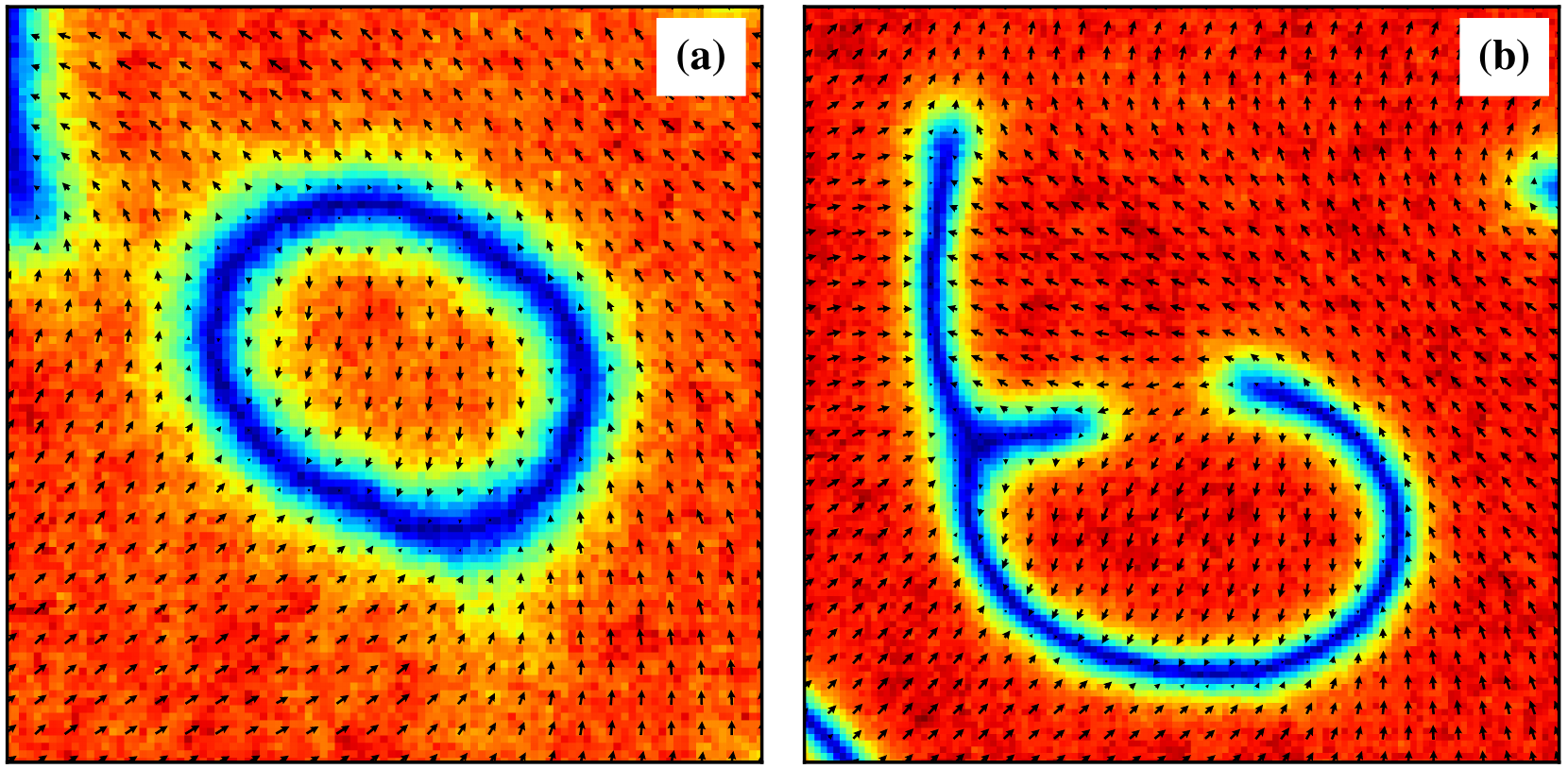}
    \caption{Panels (a) and (b) present snapshots of the $\boldsymbol{P}$-field, highlighting a closed-loop configuration and branched strings, respectively. The colormap represents the magnitude of $\boldsymbol{P}$, while the arrows indicate its local orientation. Panel (a) corresponds to a zoomed in view of panel (h) in Fig.~\ref{phase_diagram}, and panel (b) is a zoomed in view of panel (p) in the same figure.}
    \label{brunched_strings}
\end{figure*}

%%%%%%....System at high coupling and high density
\section{Effect of higher coupling strength and high density}\label{app:hdhc}
We also checked the results for high coupling $\gamma =1.5$ and density $\overline{\rho}_{p_0} = 0.8$. In FIG.\ref{high_density_high_coup_snaps} we show the local $\boldsymbol{Q}$ and $\boldsymbol{P}$ fields. Color bar and lines have the same meaning as in FIG.1 in the main text. For higher coupling the magnitude of local order parameters go beyond $1$ due to the additional ordering effect of coupling $\gamma$. The animation of the same system can be found at  Sec.\ref{movie} in the suppleimentary material.

\begin{figure*}[hbt]
    \centering
    \includegraphics[width=0.96\textwidth]{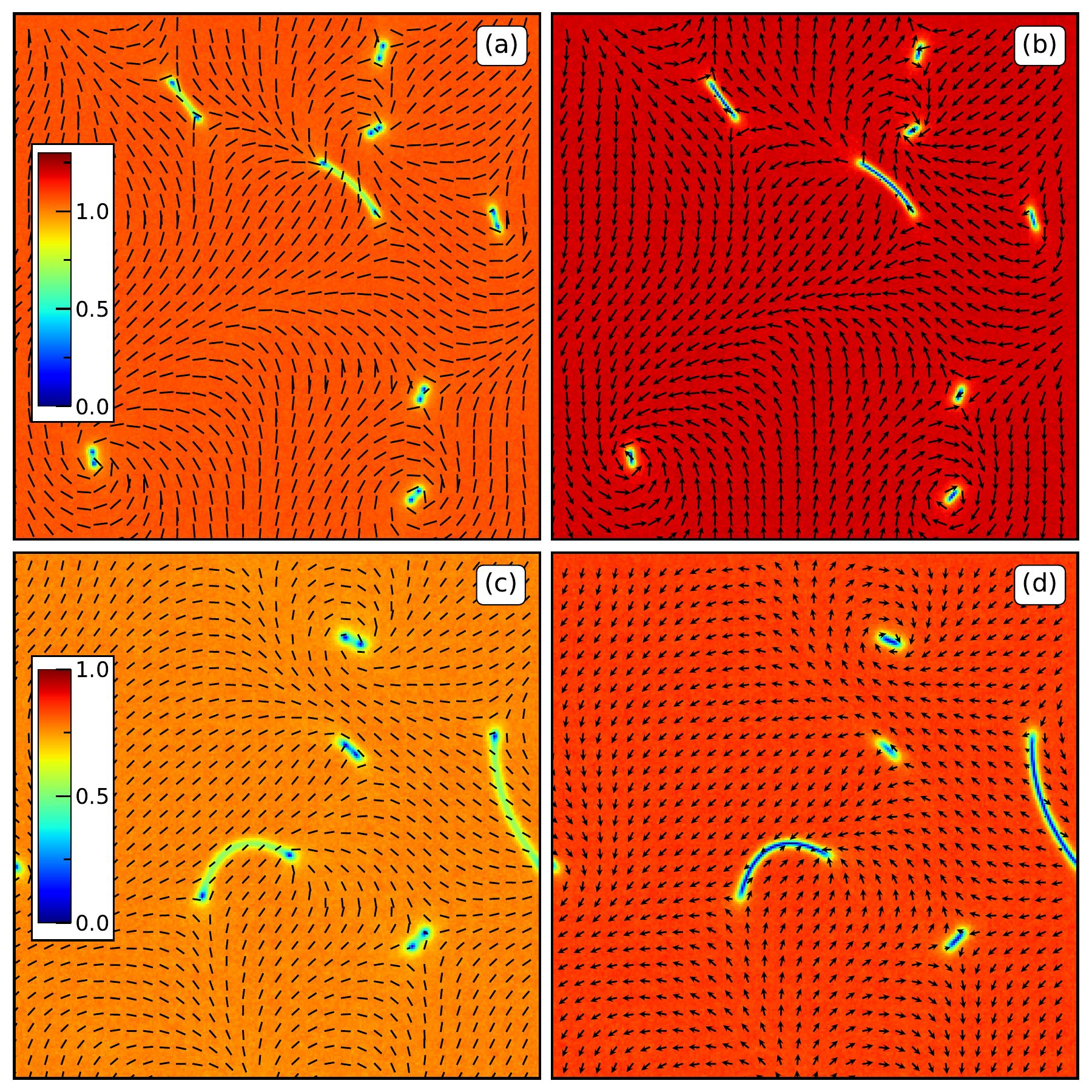}
    \caption{The figure shows the snapshots of the system at high coupling in subplots (a-b) for $\gamma = 1.50,   
     \overline{\rho}_{p_0} = 0.40$ and at high density in subplots (c-d) for $\gamma = 1.00, \overline{\rho}_{p_0} = 0.80$. Left column : snapshots of the apolar species where the color shows the magnitude of nematic order parameter, $\lvert \boldsymbol{Q} \rvert$, according to the color bar and the lines represent the orientation of the local director of the apolar species. Right column : snapshots of the polar species where the color shows the magnitude of polar order parameter, $\lvert \boldsymbol{P} \rvert$, according to the color bar and the arrows represent the local direction of the $\boldsymbol{P}$-field of the polar species. Parameters : System size,  $L = 256$, and rest of the parameters are same as in FIG.1 in the main text.}
    \label{high_density_high_coup_snaps}
\end{figure*}

%%%%%%....Kinetics of apolar order parameter   
\section{Kinetics of apolar order parameter}\label{Kinetics of apolar order parameter}
We define the two-point correlation function $C(r, t)$ = $\langle \boldsymbol{Q}({\bf r} + {\bf r_0},t) : \boldsymbol{Q}({\bf r_0}, t)\rangle$ , where $\langle \cdots\rangle$ denotes the isotropically averaged over different directions, and 50 independent runs. The system size taken into consideration is $L = 1024$. The Fig.\ref{correlation}(a-b) show the scaled correlation functions $C(r)$ vs. $r/L(t)$ for coupling strength $\gamma = 0.5, 1.0$ in sequence (main figure). The inset plots in (a-b) depict the time progression of the correlation functions $C(r)$ vs. $r$ shown in different color from black ($t = 125$) to brown ($t = 2500$).  For scaling, the characteristic length $L(t)$ is determined as the position where the correlation crosses 0.5.   
The correlation functions $C(r,t)$ display a good dynamical scaling for $\gamma = 0$ as well as 1.0. That implies the existence of single characteristic length scale $L(t)$ for the evolution for zero and finite coupling as well. But, surprisingly for intermediate coupling strength $\gamma$, we do not observe such scaling collapse (data not shown), That indicates the presence of multiple length scales in the system. 

\begin{figure*}[hbt]
    \centering
    \includegraphics[width=\textwidth]{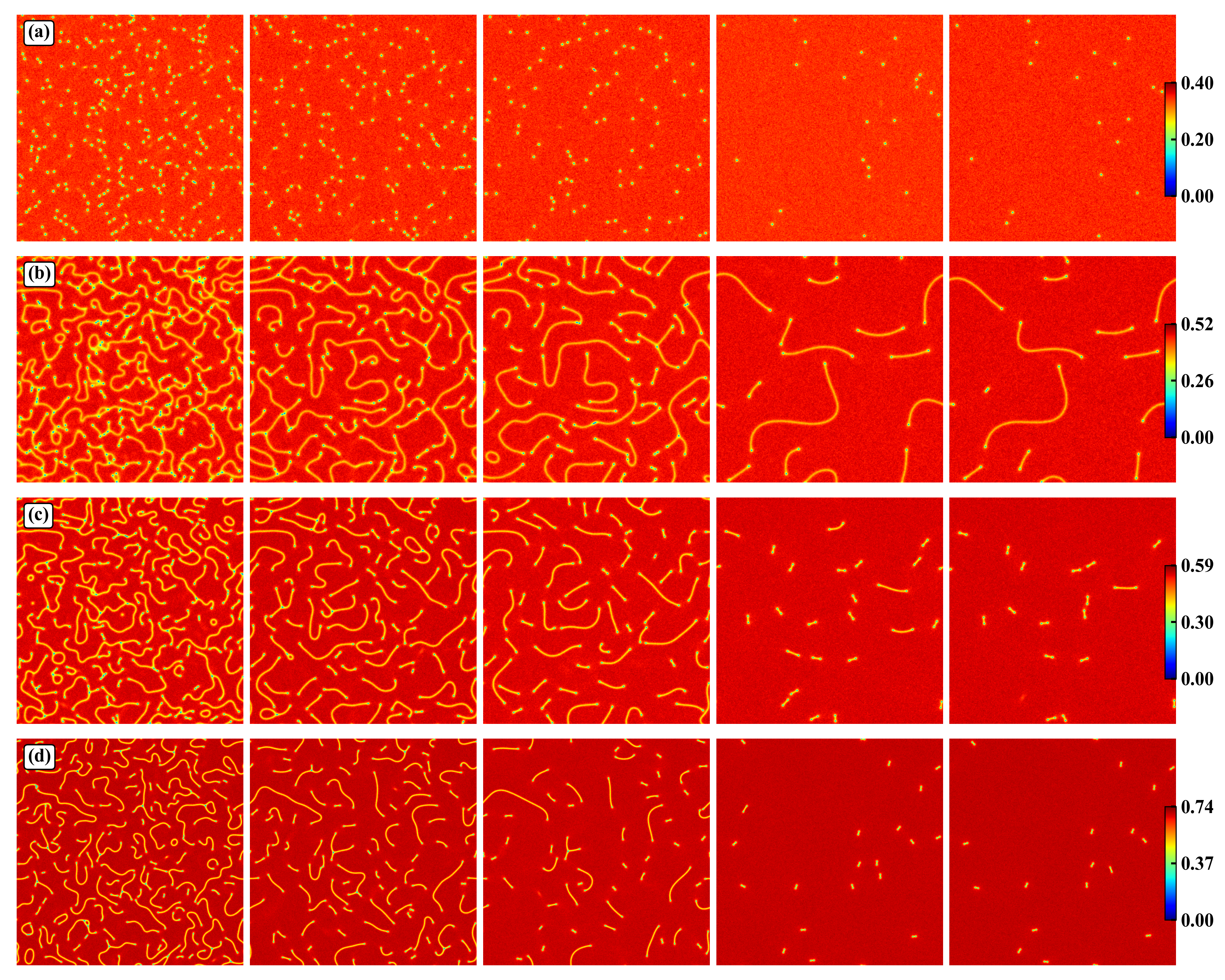}
    \caption{The snapshots illustrate the time evolution of the \({\bf  Q} \)- field across four panels (a-d), corresponding to coupling strengths \( \gamma = 0, 0.7, 0.8, 1 \), respectively. Each panel displays multiple figures, arranged from left to right, representing the evolution of \({\bf Q} \)- field over time at $t = 10, 30, 305, 650, 1200, 1500$. The heatmap in each panel indicates the magnitude of the nematic order parameter for the corresponding system at that time.}
    \label{kinetic_snap}
\end{figure*}

\begin{figure}[hbt]
%  \centering
  \includegraphics[width=0.75\textwidth]{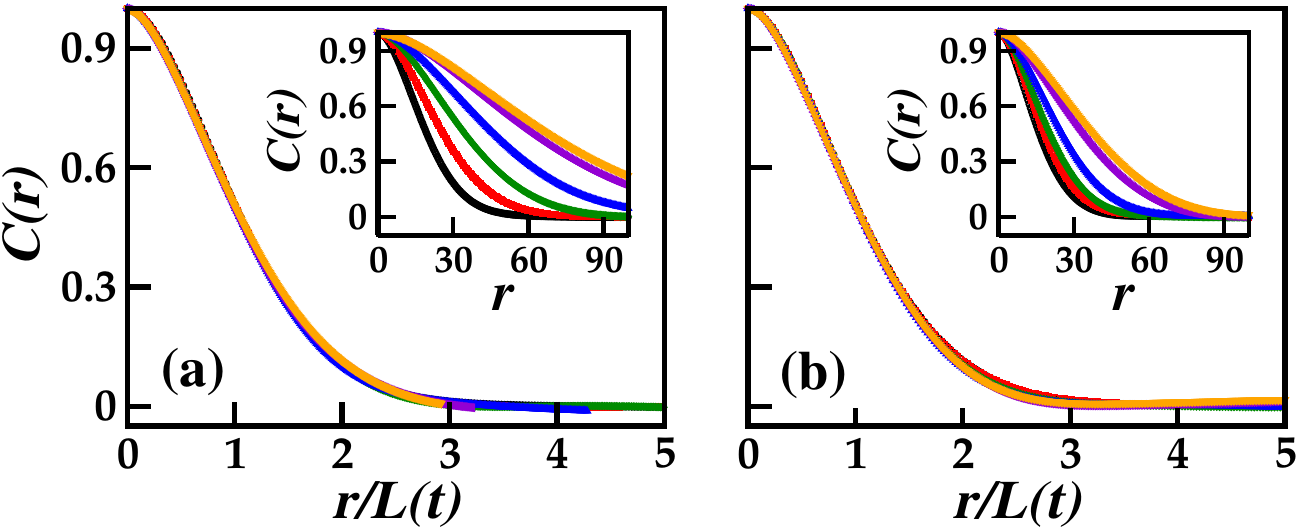}
   \caption{Panels (a) and (b) display the scaled correlation functions \( C(r) \) versus \( r/L(t) \) for coupling strengths \( \gamma = 0.5 \) and \( \gamma = 1.0 \), respectively for $\overline{\rho}_{p_0} = 0.1$ in the main figures. The curves in both the main and inset figures represent the correlation functions at different time steps, with the color progression from black ($t = 125$) to brown ($t = 2500$) indicating the time evolution. The insets in panels (a) and (b) show the correlation functions \( C(r) \) as a function of \( r \).}
    \label{correlation}
\end{figure}

%%%%%%......Artificial defects in the system
\section{Artificial Defects in the Coupled System}\label{Artificial Defects in Coupled System}
The artificial defects are designed as: Let \((x_1,y_1)\) and \((x_2,y_2)\) are the location of defect cores, then the orientation of the nematic particles at position \((x,y)\) is given by 
\begin{equation}
    \theta(\vec{r})=\Bigg[S \arctan(\frac{y-y_1}{x-x_1})-S \arctan(\frac{y-y_2}{x-x_2})\Bigg]+\theta_0
\end{equation}
where, \(\theta_0=0\) is the orientation far away from the defects such that nematic particles are aligned parallel to the horizontal axis and \({S=\frac{1}{2}}\) is the topological charge of the defect \cite{artificial_defect}. 
 We chose \((x_1,y_1)\equiv(\frac{L}{4},\frac{L}{2})\) and \((x_2,y_2)\equiv(\frac{3L}{4},\frac{L}{2})\) and the orientation of the defects are chosen such that the head of the \(+\frac{1}{2}\) defect is facing the \(-\frac{1}{2}\) as shown in FIG. \ref{fig_artificial}.

Snapshots of the system with polar order initial condition are shown for \(\gamma = 0\) and \(\gamma = 1.0\). These snapshots provide insight into the initial configurations and the subsequent evolution of the system under different \(\gamma\) values.

\FloatBarrier
\begin{figure*}[hbt]
    \includegraphics[width=\textwidth]{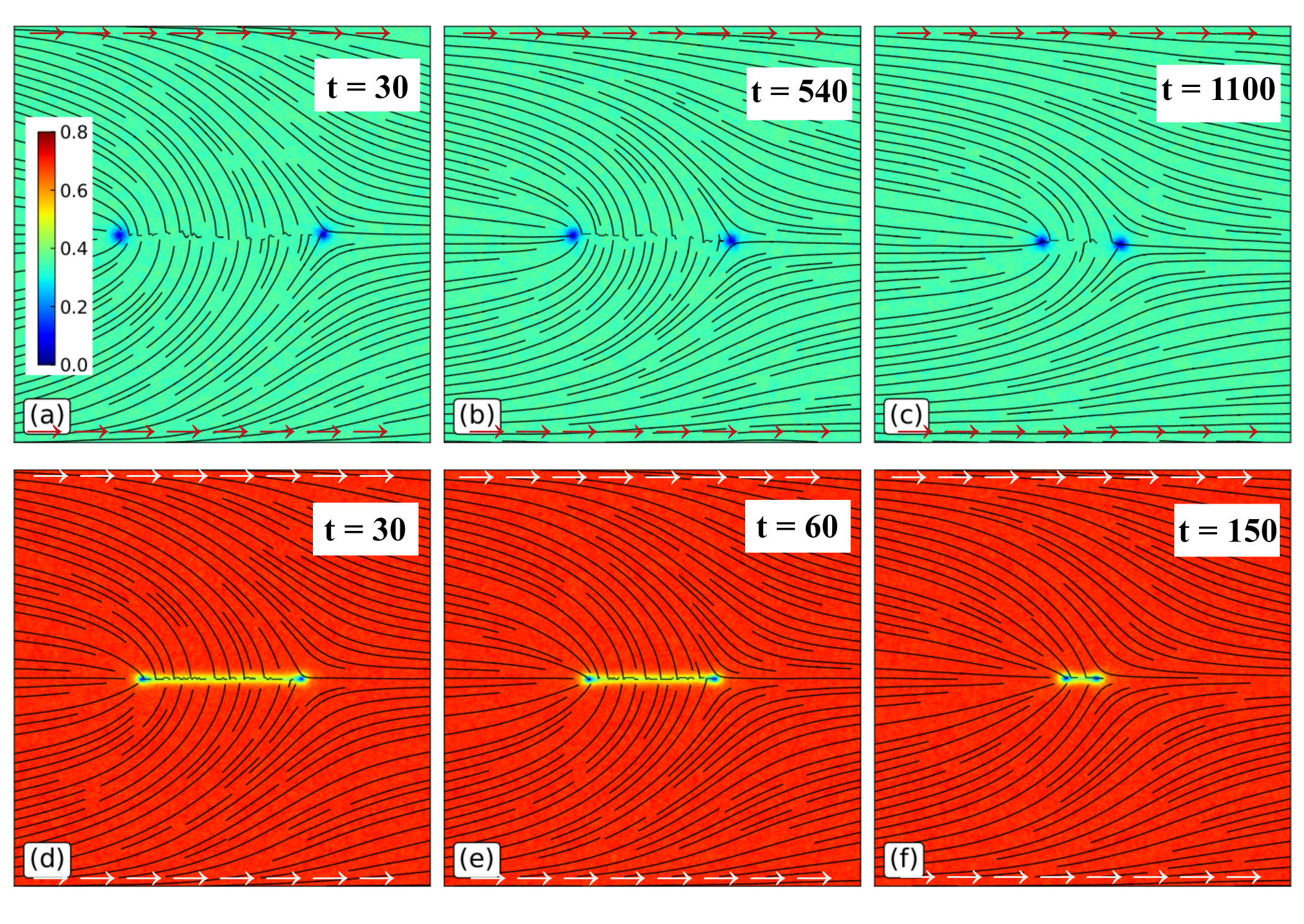}
    \caption{Snapshots of the $\boldsymbol{Q}$-field with artificial defects. (a-c) For \(\gamma = 0\), snapshots are displayed at time \(t=\) \(30\), \(540\), and \(1100\), respectively. (d-f) For \(\gamma = 1.0\), snapshots are taken at time \(t=\) \(30\), \(60\), and \(150\), respectively. The color map represents the magnitude of the $\boldsymbol{Q}$-field, streamlines depict its orientation, and arrows along the boundary indicate the $\boldsymbol{P}$-field.}
    \label{fig_artificial}
\end{figure*}
\FloatBarrier

% \newpagep

\section{Description of Supplementary Movies}\label{movie}
\textbf{Mov1 :} The movie shows the evolution of the nematic order parameter field ($\boldsymbol{Q}$) for $\gamma = 0.0$, $\overline{\rho}_{p_0}$ $= 0.40$. The colormap shows the magnitude of $\boldsymbol{Q}$ and the headless arrows represent the orientation of the local nematic director.\\
{\em Link} : \url{https://drive.google.com/file/d/1r7Y5IeG9q8QwlL7jP5tIotpX1J_cedWw/view?usp=sharing}\\

\textbf{Mov2 :} The movie shows the evolution of the nematic order parameter field ($\boldsymbol{Q}$) for $\gamma = 0.70$, $\overline{\rho}_{p_0}$ $= 0.40$. The colormap shows the magnitude of $\boldsymbol{Q}$ and the headless arrows represent the orientation of the local nematic director.\\
{\em Link} : \url{https://drive.google.com/file/d/1Lt3JFlml3pw957CAeuoElqJ-Pt338d4i/view?usp=sharing}\\

\textbf{Mov3 :} The movie shows the evolution of the polar order parameter ($\boldsymbol{P}$) for $\gamma = 0.70$, $\overline{\rho}_{p_0}$ = $0.40$. The colormap shows the local magnitude of $\boldsymbol{P}$ and the arrows represent the local orientation of $\boldsymbol{P}$.\\
{\em Link} : \url{https://drive.google.com/file/d/1rK_UUUuAxk2LqWEI_eLjRTo5QiRQsQJd/view?usp=sharing}\\

\textbf{Mov4 :} The movie shows the evolution of the nematic order parameter field at high mean density of polar species ($\overline{\rho}_{p0} = 0.60$) for $\gamma = 1$. The colormap shows the local magnitude of $\boldsymbol{P}$ and the arrows represent the local orientation of $\boldsymbol{P}$.\\
{\em Link} : \url{https://drive.google.com/file/d/1a8V6Pe_c_g9pHWA3qJKujTsSygyFMVNb/view?usp=sharing}\\

\twocolumngrid
 
\nocite{*}
%\newpage
\bibliography{apssamp}% Produces the bibliography via BibTeX.

\end{document}